\documentclass[namedreferences]{kluwer}    % Specifies the document style.

\usepackage{epsfig}

\begin{document}  
\begin{article}
\begin{opening}         

\title{Attentional modulation of firing rate and synchrony in a
            model cortical network} 
\author{Buia \surname{Calin}\email{buia@physics.unc.edu}}  
\author{Paul \surname{Tiesinga}\email{tiesinga@physics.unc.edu}}  
\runningauthor{}
\runningtitle{Attentional modulation of firing rate and synchrony in a
            model cortical network}
\institute{Dept. of Physics and Astronomy, University of North
            Carolina at Chapel Hill}
\date{\today}

\begin{abstract}
The response of a neuron in the visual cortex to stimuli of different
contrast placed in its receptive field is commonly characterized using
the contrast response curve. When attention is directed into the
receptive field of a V4 neuron, its contrast response curve is shifted to
lower contrast values (Reynolds et al, 2000, Neuron 26:703). The neuron
will thus be able to respond to weaker stimuli than it responded to
without attention. Attention also increases the coherence between neurons
responding to the same stimulus (Fries et al, 2001, Science 291:1560).
We studied how the firing rate and synchrony of a densely interconnected
cortical network varied with contrast and how they were modulated by
attention. The changes in contrast and attention were modeled as changes
in driving current to the network neurons.
We found that an increased driving current to the excitatory neurons
increased the overall firing rate of the network, whereas variation of
the driving current to inhibitory neurons modulated the synchrony of the
network. We explain the synchrony modulation in terms of a locking
phenomenon during which the ratio of excitatory to inhibitory firing
rates is approximately constant for a range of driving current values.
We explored the hypothesis that contrast is represented primarily as a
drive to the excitatory neurons, whereas attention corresponds to a
reduction in driving current to the inhibitory neurons.  Using this
hypothesis, the model reproduces the following experimental observations:
(1) the firing rate of the excitatory neurons increases with contrast;
(2) for high contrast stimuli, the firing rate saturates and the network
synchronizes; (3) attention shifts the contrast response curve to lower
contrast values; (4) attention leads to stronger synchronization that
starts at a lower value of the contrast compared with the attend-away
condition. In addition, it predicts that attention increases the delay
between the inhibitory and excitatory synchronous volleys produced by the
network, allowing the stimulus to recruit more downstream neurons.
\end{abstract}

\keywords{attention, synchrony, V4 area}

\end{opening}           

\section{Introduction}

Neurons in cat primary visual cortex are orientation selective as they
respond with maximal firing rates to a bar of their preferred orientation
\cite{hu59,huwi59,huwi62}. The firing rate increases
with stimulus contrast \cite{sclar82}. In addition, when
attention is directed into the receptive field of V2 and V4 neurons, the
firing rate in response to a visual stimulus inside the receptive field
increases \cite{luck97,mcad99}. Orientation
selectivity is preserved across different values for the contrast and
attentional state. A fundamental question of neuroscience is how this
invariant tuning is achieved and maintained in local cortical circuits.
This requires a more detailed understanding of how the firing rate in
cortical circuits is regulated by attention and contrast than is
presently available.

The contrast response function (CRF, the neuron's firing rate plotted
as a function of luminance contrast) has a 
characteristic sigmoidal shape \cite{albr82,ohz82,sclar82,albr02}.  In a study
of the attentional modulation of the CRF of V4 neurons it was found that
the sensitivity of neurons was increased (contrast gain modulation),
because the CRF was shifted to the left \cite{rey00,rey03}. Thus, the
neurons were activated by low contrast 
stimuli to which they did not respond when attention was directed away
from the neuron's receptive field.  For high enough contrast, the
local field potential 
had a broad peak in the gamma-frequency range (30-80 Hz)
\cite{gray97,hen05}, suggesting that the cortical
network in which the neuron is embedded oscillates in the gamma-frequency
range. Likewise, when stimuli were presented that drove the cell
strongly, attention increased the correlations between cells responding
to similar stimuli and only weakly increased the neuron's firing rate
\cite{fries01,bich05}. These results 
provide support for the idea that the effects of attention on firing rate
and neural synchrony are equivalent to changing the effective contrast of
the stimulus. Despite these similarities, it is not clear whether
attention and contrast use the same mechanism to modulate the neuron's
responses.  Nor is it clear whether contrast and attention act 
similarly on the correlations between neurons.

The response of a single neuron model to temporally patterned excitatory
and inhibitory synaptic inputs was studied previously \cite{ties04nco}.
Attention was postulated to alter the
precision of presynaptic spike trains \cite{ties04nco}. The model
did not offer a mechanistic explanation for how these changes could
occur. Here, we study attentional modulation of a strongly coupled model
network, representative of those in the superficial layers of cortex
\cite{yoshi05}. In cat and macaque primary visual cortex, the
stimulus orientation that most strongly activates neurons varies
systematically with the location on the cortical surface
\cite{blas92a,blas92b}. Similar, but more elaborate maps are also
thought to exist in 
downstream cortical areas such as V2 and V4 \cite{gho97}. These
maps have not been studied in great detail. Here we study the response of
neurons at one particular cortical location to a stimulus of the optimal
orientation.  The basic assumption is that attention and changes in
contrast are mediated by changes in driving currents to the excitatory
and inhibitory neurons in the network.  Our simulation results reveal
that increasing the drive to excitatory neurons has a different effect
than decreasing the drive to inhibitory neurons. Specifically, varying
the driving current to the interneurons leads to synchrony modulations by
way of so called locking steps. We explore whether it is possible to
account for the experimental results under the hypothesis that contrast
drives the excitatory neurons more strongly than the inhibitory neurons
and that attention leads to a reduction in the drive to the inhibitory
neurons.

\section{Methods and model description}

\subsection{Model summary}

\begin{figure}
\centering
\epsfig{file=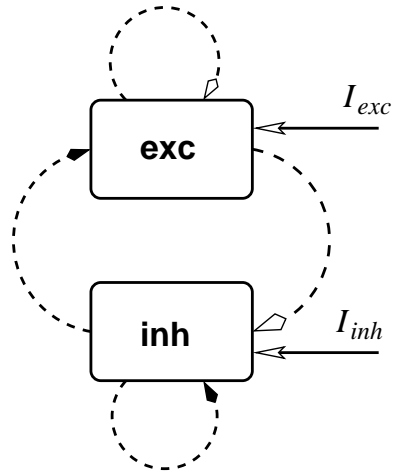, width=2in}
\caption{The model network consisted of a fully connected set of
excitatory and inhibitory neurons. Each excitatory and inhibitory neuron
also received a constant driving current equal to $I_{exc}$ and
$I_{inh}$, respectively.} 
\label{f:model}
\end{figure}

We use a generic network model that is an approximation for local
networks in V1, V2 and V4.  We focus our attention on a subnetwork of
strongly interconnected neurons, based on observations made in layer 2/3
of rat primary visual cortex made by \cite{yoshi05}. The model
consisted of four hundred excitatory pyramidal neurons and a hundred
inhibitory interneurons (Figure \ref{f:model}). In the model used for
the initial 
exploration, the neurons were connected all-to-all with inhibitory
GABA$_\textrm{A}$ 
synapses and excitatory AMPA synapses. For the robustness part of the
study a sparsely coupled network was used. In addition to the excitatory
and inhibitory inputs from other neurons in the subnetwork, the neurons
also received feed-forward stimulus-related excitatory inputs
(``contrast''), modulatory top-down inputs (``attention'') and inputs
from other neurons in the same layer, which are not part of 
the simulated network. All of these inputs were modeled as a time varying
current, the mean of which was $I_{exc}$ ($I_{inh}$) and the variance
of which was $\lambda_{exc}$ ($\lambda_{inh}$) for excitatory
(inhibitory) cells. We used Hodgkin-Huxley-type 
model neurons for excitatory pyramidal cells (Golomb and Amitai, 1997)
and for inhibitory interneurons (Wang and Buzsaki, 1996). A full
description of the single neuron model equations, the synapse models, and
details regarding computational implementation are given in the appendix.
In summary, the key parameters varied in the simulation were: the driving
current $I_{exc}$ ($I_{inh}$), the current-noise strength
$\lambda_{exc}$ ($\lambda_{inh}$), and the heterogeneity parameters
$\sigma_{exc}$ ($\sigma_{inh}$) for excitatory (inhibitory) 
neurons as well as the synaptic coupling parameters ($g_{ee}$,
$g_{ei}$, $g_{ie}$, $g_{ii}$), defined as the unitary synaptic
strength times the number of inputs to the receiving neuron. The subscript
``$ie$'' stands for the inhibitory input to the excitatory neurons, with an
analogous interpretation for ``$ii$'', ``$ee$'' and ``$ei$''. Driving
currents and heterogeneity are expressed in $\mu \textrm{A/cm}^2$,
noise strength  in $\textrm{mV}^2/\textrm{ms}$ and synaptic strength in $\textrm{mS/cm}^2$.

We performed many exploratory simulations. For almost all combinations of
synaptic coupling parameters that were studied, we found values for the
excitatory and inhibitory driving current for which the network was in a
synchronous state. Not all of these states were equally robust against
noise and heterogeneity. \cite{borg03,borg05,borg05b} report on
procedures for finding coupling parameters that yield robust
synchronization. The authors suggest that the loss of synchrony proceeds
either by ``phase walkthrough'', for which the interneurons receive
enough driving current and tonic excitatory synaptic inputs to spike
without having to wait for synchronous excitatory inputs, or via
``suppression'', for which the interneurons fire at such a high rate
or at such a low degree of synchrony that they prevent the excitatory
neurons from firing. 
They also note that asynchronous interneurons are more effective in
suppressing excitatory cells than synchronous interneurons firing at the
same rate. To obtain robust synchrony \cite{borg03,borg05,borg05b}:
(1) the current drive to the interneurons should be low enough that
synchronous excitatory synaptic inputs are necessary for spiking, in
order to prevent phase walkthrough; (2) the ratio of the inhibitory
synaptic current to excitatory neurons over the excitatory synaptic
current should be large enough to ensure robustness, but small enough to
prevent suppression. We picked for further analysis the following set of
coupling parameters ($g_{ee}$, $g_{ei}$, $g_{ie}$, $g_{ii}$)= (0.05,
0.15, 0.15, 0.15) consistent with these requirements. The parameter
($\lambda_{exc}$, $\lambda_{inh}$) took two values: (0.02, 0.1) and
(0.2, 0.6), corresponding to low and high noise states,
respectively. The heterogeneity parameters were $\sigma_{exc}=0.1$ and
$\sigma_{inh}=0.02$. 

To test the robustness of our findings for the all-to-all network, we
performed simulations of a network with sparse synaptic connections, high
noise levels and large heterogeneities in the drive to the excitatory
cells. All connections were made stochastically with a probability of 20\%
\cite{bru03}, with the exception of the recurrent inhibitory
connections, which remained all-to-all. The synaptic coupling strengths
were made stronger, $g_{ee}=0.1$ (80), $g_{ei}=0.5$ (80), $g_{ie}=0.3$
(20) and $g_{ii}=0.5$ (100). The number between the parentheses
indicates the average number of 
connections the neuron received. A hundred excitatory cells (referred to
as the top-100) received a driving current $I_{exc}$ with an offset varying
linearly from 0.5 for the first neuron to -0.5 for the hundredth neuron;
the remaining three hundred (referred to as the bottom-300) received a
driving current $I_{exc}-1$, also with a linear offset between -0.5 and 0.5.
Note that highest current for the bottom-300 is equal to the lowest
current for the top-100. The current to the inhibitory neurons was
normally distributed with mean $I_{inh}$ and a standard deviation of
$\sigma_{inh}=0.1$. The voltage noise variances are
$\lambda_{exc}=0.1$ and $\lambda_{inh}=0.5$. 

\subsection{Analysis}

The spike time was defined as the time at which the membrane potential
crossed a threshold value from below. The threshold value was taken to be
0 mV for interneurons and -20 mV for pyramidal neurons. The mean firing
rate was calculated as the inverse of the mean interspike interval for
each neuron, averaged over all neurons of the same type (i.e. separately
for all the inhibitory and all the excitatory neurons):

\[r=\frac{1}{N}\sum_i \left( \frac{1}{n^i-1} \sum_j(t^i_{j+1}-t^i_j)
\right)^{-1} 
\]
where $t_j^i$ was the $j^{th}$ spike time of neuron $i$, $n_i$ was the
number of spikes produced by neuron $i$, and $N$ was the total number
of neurons of a given type. In this notation, $r_{exc}$ and $r_{inh}$
were the average firing rates of the excitatory and the inhibitory
neurons, respectively. 
The degree of synchrony of the system was estimated using the coefficient
of variation for a population of neurons, $CV_P$. This measure of synchrony
is based on the idea that during synchronous states the minimum distance
between spikes of different neurons is reduced compared with asynchronous
states. First, the spike times of the whole population of neurons were
pooled together and sorted in ascending order. The sorted set of
aggregate spike times ${t_j}$ was labeled by the index $j$. The interspike
interval between two consecutive spikes
$t_j=t_{j+1}-t_j$ typically involved spikes from two different neurons.  The
coefficient of variation of the combined interspike intervals is:

\[ CV=\frac{\sqrt{\langle \Delta t_j^2 \rangle - \langle \Delta t_j
    \rangle ^2}}{\langle \Delta t_j \rangle}
\]
Here $\langle \bullet \rangle$  was the average across all intervals
$j$.  The value of the $CV$ is 
one for an asynchronous network \cite{ties04}, whereas
for a perfectly synchronous network it is approximately $\sqrt{N}$
\cite{ties04}, with $N$ being the number of active
neurons in the network. In the model, there were four times as many
excitatory neurons 
as there were inhibitory neurons. Hence, in order to compare the
synchrony of the excitatory neurons to that of the inhibitory neurons, we
subtract one from the $CV$, and, in addition, divide it by
$\sqrt{N_{exc}/N_{inh}}=2$ for the 
excitatory neurons. The resulting quantity is denoted by $CV_P$ (this is
different from the normalization used previously in \cite{ties04}). A
$CV_P$ value close to zero then corresponds to an 
asynchronous state, while a value greater than zero indicates synchrony;
the higher the synchrony, the higher the $CV_P$ is. There are two different
aspects of synchrony: the degree of coincidence and the level of
precision. The $CV_P$ confounds coincidence and precision
\cite{ties04}. Specifically, for a synchronous network in which the 
neurons fire only once every few cycles, but when they do so at high
precision, the $CV_P$ can have a value comparable to that of a network where
neurons fire every cycle but at a lower precision.
The delay between inhibitory and excitatory volleys was determined based
on the interspike interval (ISI) distribution. For each interneuron
spike, we first determined all the ISIs with a given excitatory neuron.
We only keep the ISI with the smallest absolute value and then repeat
this procedure for all excitatory neurons and across all interneuron
spikes. This yields $N_{inh}^{sp}\times N_{exc}$ interspike intervals,
where $N_{inh}^{sp}$ was the total number of inhibitory spikes and
$N_{exc}$ was the number of excitatory neurons. The 
distribution of ISIs had at least one peak when the network was
synchronized. The location of the peak closest to zero ISI was taken as
an estimate of the delay between the excitatory and inhibitory spike
volleys. This delay could be positive or negative.
Spike time histograms ($STH$) were obtained as

\[STH=\frac{1000 \Delta t}{N}\sum_{i=1}^N X_i(t)
\]
where

\[X_i(t)=\left\{ \begin{array}{lc} 1, & \textrm{if } t^i_j \in
  (t-\Delta t/2,t+\Delta t/2) \\
0, & \textrm{otherwise} \end{array}
\right.
\]
$N=N_{exc}$ or $N_{inh}$ was the number of neurons of a given type and
the bin width was $\Delta t=2$ ms.

The frequency, $f_{osc}$, of the network oscillation was determined as the
location of the highest peak in the Fourier transform of the interneuron
spike time histogram.

\section{Results}

\begin{figure}
\centering
\epsfig{file=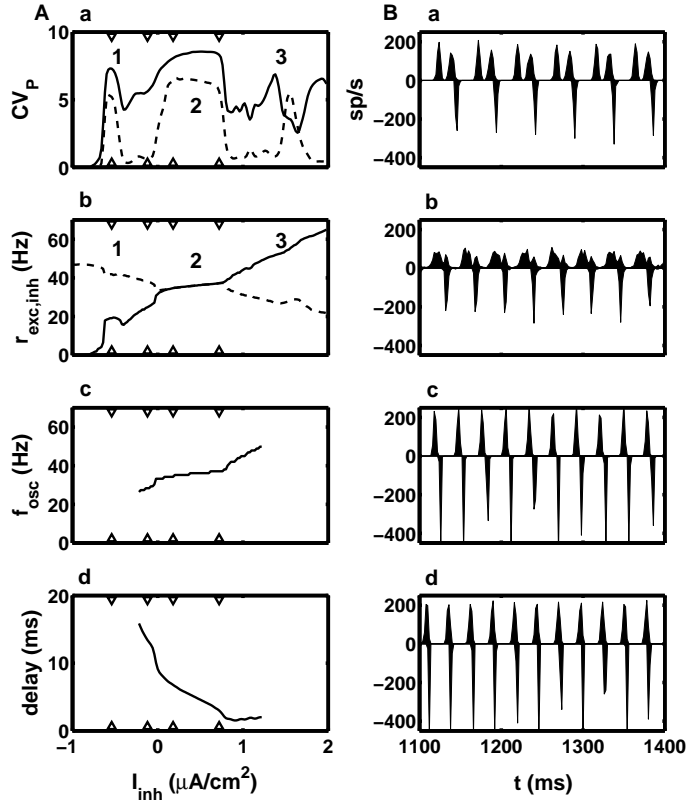, width=4in}
\caption{Network synchrony and the delay between inhibitory and
excitatory volleys were modulated by the inhibitory driving current. (A)
We plot as a function of the inhibitory driving current $I_{inh}$, (a)
$CV_P$ for 
the excitatory (dashed line) and the inhibitory neurons (solid line), (b)
the firing rate of the excitatory (dashed line) and inhibitory neurons
(solid line), (c) the oscillation frequency when the network is in or
near a 1:1 locking state, (d) the delay between the inhibitory volley and
the excitatory volley. (B) In each panel, we plot the spike time
histogram for the excitatory neurons (positive rates) and inhibitory
neurons (negative rates). The driving currents used in (a) to (d)
correspond to the currents indicated by the open triangles in A, from
left to right, respectively. The numbers in Aa and Ab are described in
the main text. The driving current to the excitatory neurons was
$I_{exc}=3.94 \, \mu\textrm{A/cm}^2$ and we used a low noise level.}
\label{f:inhidep}
\end{figure}

Our goal is to characterize first how synchrony and firing rate are
modulated in cortical networks (Figure \ref{f:inhidep} to
\ref{f:sensmod}), and then link these 
simulation results to the observed effects of attention on the contrast
response function (Figure \ref{f:ispace} to \ref{f:delaymod}). When
the driving current to the 
excitatory cells is increased, their firing rate increases. Recurrent
excitatory connections cause a larger excitatory drive to both the
excitatory neurons as well as the inhibitory neurons. Hence, the
excitatory cells also receive more inhibition. The net increase in the
firing rate of the excitatory and the inhibitory neurons will depend on
how much inhibition is recruited, which is proportional to both $g_{ei}$ and
$g_{ie}$. What would happen if the driving current to the inhibitory cells was
increased instead? The rate of the inhibitory neurons would initially
increase, but that increase would reduce the firing rate of the
excitatory neurons, and the rate of excitatory inputs to the interneurons
as well. The net increase in the interneuron firing rate and the decrease
in the firing rate of the excitatory neurons again depends on the
relative strength of $g_{ei}$ and $g_{ie}$. The preceding analysis is
based only on 
the changes in firing rate. However, there may also be changes in
synchrony that could lead to an increase or a reduction in the
effectiveness of the synaptic inputs in making the postsynaptic cells
spike \cite{sal00,ties04p}.

\subsection{Variation of the drive to the inhibitory neurons.}

Synchrony modulations lead to so called $n:m$ locking steps, during which $n$
synchronous excitatory volleys are followed by $m$ synchronous inhibitory
volleys (here either $n$ or $m$ is one). In Figure \ref{f:inhidep} we show
the behavior of 
the all-to-all network at a low noise level. Three locking steps are
visible, the 2:1, 1:1 and 1:2 steps, labeled by 1, 2 and 3, respectively
(Figure \ref{f:inhidep}Aa and  \ref{f:inhidep}Ab). On the locking
steps the network was highly 
synchronous, as quantified by the $CV_P$ (Figure  \ref{f:inhidep}Aa), and
the excitatory 
and inhibitory firing rates varied little with $I_{inh}$ (Figure
\ref{f:inhidep}Ab). On the 
1:1 step, the excitatory and inhibitory firing rates were equal to the
oscillation frequency (Figure  \ref{f:inhidep}Ac). In addition, the
delay between 
inhibitory and excitatory volleys decreased monotonically from 6.78 ms at
the lowest current value on the step to 3.01 ms at the highest current
value on the step. The histograms of the corresponding network activity
are shown in Figure  \ref{f:inhidep}Bb and  \ref{f:inhidep}Bc,
respectively. A reduction in synchrony 
by way of the phase walkthrough mechanism (Borgers and Kopell, 2005) was
signaled by the delay approaching zero. During phase walkthrough the
interneurons receive enough driving current to be able to spike without
having to wait for synchronous excitatory volleys. The reduction in
synchrony induced by increasing $I_{inh}$ on the 1:1 step led to an increase
in inhibitory firing rate, whereas on the 2:1 step it led to a decrease
(Figure  \ref{f:inhidep}Ab). The former is expected in view of the
increased inhibitory driving current, the latter is the
exception. \footnote{This comes about as follows. For $I_{inh}$ values
  below the onset of the 2:1 
step, the asynchronous excitatory synaptic inputs to the interneurons
were below threshold. Only synchronized excitatory volleys, with
approximately the same number of excitatory spikes, could elicit spikes
from the interneurons. Hence, the onset of the 2:1 step was associated
with a sharp increase in inhibitory rate. In contrast, when the step
became unstable, the asynchronous excitation was less efficient in
driving the interneurons than the synchronous volleys were on the locking
step, yielding a sharp reduction in firing rate.
}

\begin{figure}
\centering
\epsfig{file=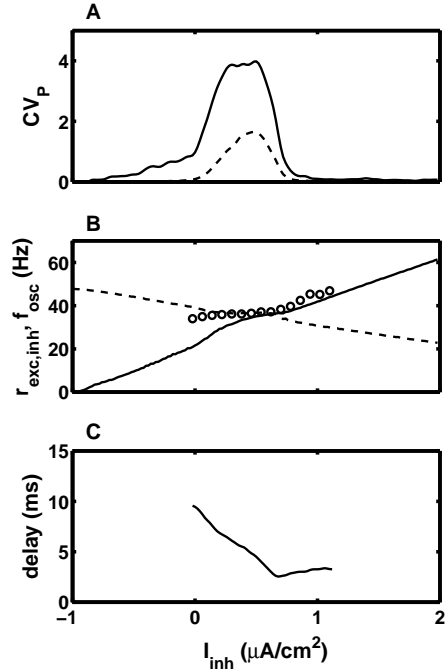, width=2.5in}
\caption{For high noise levels, synchrony only occurred for a limited
range of driving currents. We plot, (a) $CV_P$ for excitatory (dashed line)
and inhibitory neurons (solid line), (b) the firing rate of excitatory
(dashed line) and inhibitory neurons (solid line) and the oscillation
frequency (open circles), (c) the delay between inhibition  and
excitation. The quantities are plotted as a function of the inhibitory
driving current $I_{inh}$. The simulations were performed for a high noise
level with $I_{exc}=3.94 \, \mu\textrm{A/cm}^2$.}
\label{f:hnoise}
\end{figure}

 In Figure \ref{f:hnoise} we show the behavior of the same network but
 at high noise levels. Only the 1:1 step remained. It was
 characterized by a peak in the 
$CV_P$ (Figure \ref{f:hnoise}A), but, in contrast to the results for a
 low noise level, 
the excitatory and inhibitory rate were not exactly equal because the
interneuron skipped oscillation cycles (Figure
 \ref{f:hnoise}B). However, as before, 
the delay between inhibition and excitation decreased systematically with
$I_{inh}$ (Figure \ref{f:hnoise}C). Outside the 1:1 step the network
 is asynchronous, with 
$CV_P$ values close to zero (Figure \ref{f:hnoise}A).  These results
 show that a low noise strength is not a necessary condition for the
 presence of 1:1 locking states and modulation of synchrony and delay
 with interneuron drive. 

In summary, due to the presence of locking states, the rate of increase
of $r_{inh}$ and the rate of decrease of $r_{exc}$ with inhibitory
driving current varied non-monotonously, and large modulations in the
degree of synchrony were obtained.

\begin{figure}
\centering
\epsfig{file=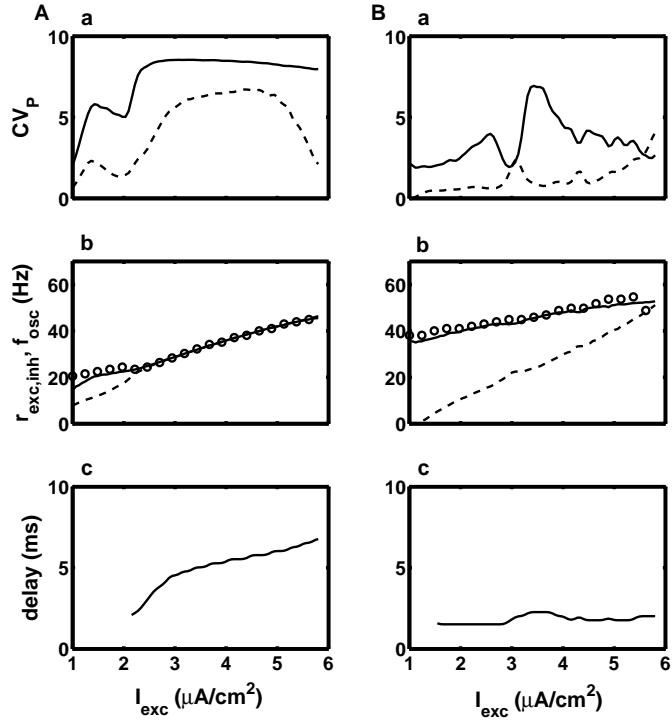, width=4in}
\caption{(A) For a small depolarizing current to the interneuron
network, synchronous states were obtained for a wide range of excitatory
driving currents. We plot as a function of the excitatory driving current
$I_{exc}$, (a) $CV_P$ for the excitatory (dashed line) and the
inhibitory neurons 
(solid line), (b) the firing rate of the excitatory (dashed line) and the
inhibitory neurons (solid line) and the oscillation frequency (circles),
(c) the delay between inhibition and excitation. (B) For a high
depolarizing current to the interneuron network, synchrony resonances
were observed. Panels are as in (A). The driving current to the
inhibitory neurons was $I_{inh}=0.38 \, \mu\textrm{A/cm}^2$ for (A)
and $I_{inh}=1.18 \, \mu\textrm{A/cm}^2$ for (B) and we used a low
noise level.} 
\label{f:excidep}
\end{figure}

\subsection{Variation of the drive to the excitatory neurons.}

When the drive to the inhibitory neurons was increased, the inhibitory
rate increased whereas the excitatory rate decreased. Therefore,
irrespective of what the initial firing rate of the excitatory neurons
was, there would be an inhibitory driving current for which the
inhibitory and excitatory rates would be approximately equal, so that
there could be locking steps. When the excitatory current is increased,
both the excitatory as well as inhibitory rates increase. Therefore the
two rates do not necessarily converge to each other. We investigate two
regimes based on the value of the driving current to the inhibitory
cells.

First, for a weak depolarizing drive to the interneurons ($I_{inh}=0.38$,
Figure \ref{f:excidep}A), the excitatory and inhibitory firing rates
were comparable at 
the start of the interval over which excitatory current was varied. 
Initially, the inhibitory cells fired at twice the rate of the excitatory
cells, but eventually the excitatory and inhibitory rates became equal,
at which point the network fully synchronized (Figure \ref{f:excidep}Aa).  The
synchronous state was very stable against increases in driving current,
yielding a range of firing rates from 25 to 45 Hz for which the network
was synchronized (Figure \ref{f:excidep}Ab). The delay went from about
2.26 ms at the start to 6.53 ms at the end of the locking step (Figure
\ref{f:excidep}Ac). \footnote{As mentioned before, the loss of
  synchrony proceeds via a phase 
walkthrough mechanism, signaled by the vanishing of the delay between
inhibitory and excitatory volleys. The region in the $I_{exc}-I_{inh}$
parameter 
plane for which 1:1 locking is obtained has a shape that resembles the
inside of an ellipse. The long axis is closer in direction to the $I_{exc}$
axis, whereas the short axis is more parallel to the $I_{inh}$ axis. This
means that it is easier to leave the locking step by varying $I_{inh}$
compared with varying $I_{exc}$. As a result the rate of change in delay with
$I_{inh}$ is also larger. Whether the delay on the 1:1 locking step increases
or decreases with increasing $I_{exc}$ depends on whether $I_{exc}$ moves away
from, or closer to, the high $I_{inh}$ edge of the locking region.  For the
parameter values used in Figure 4A, the delay increased.} Second, for
a strong depolarizing drive 
($I_{inh}=1.18$, Figure \ref{f:excidep}B), the 
interneurons fired at a much higher rate than the excitatory cells. The
$CV_P$ value for the excitatory neurons was less than for the interneurons
(Figure \ref{f:excidep}Ba). However, this was due to their low firing
rate: when the 
excitatory neurons fired on an oscillation cycle they did so at high
precision (see Methods).  There was a local increase in the interneuron
coherence when the oscillation frequency was around 45Hz, indicating a
preference for oscillations in the middle of the gamma-frequency range.
In this state, the excitatory neurons were mainly following the
inhibitory rhythm, smoothly increasing their rate with driving current
(Figure \ref{f:excidep}Bb). In addition, the delay was virtually
constant over the entire current range (Figure \ref{f:excidep}Bc).

Thus, in general, delay modulations are reduced when the drive to the
excitatory neurons is varied compared with when the drive to the
interneurons is varied. In addition, on the locking step, the inhibitory
and excitatory rates vary with the excitatory driving current, but their
ratio is constant.

\subsection{Robustness of locking states in sparse and heterogeneous networks}

The results presented in the preceding sections were obtained using an
all-to-all connected network with small heterogeneities in the driving
currents. Because the degree of heterogeneity was so small, we refer to
this network as the homogenous, all-to-all network. Cortical networks are
characterized by a sparse synaptic connectivity and heterogeneity in
neural properties \cite{binz04,shep05,song05}. Furthermore, during
synchronous states in vivo, neurons might 
not fire on each cycle of the oscillation \cite{fries01}. We
therefore explored whether there are locking states in sparsely coupled,
heterogeneous networks for which the mean inhibitory and excitatory
firing rate is less than the oscillation frequency and whether in these
states the delay between inhibitory and excitatory volleys could still be
modulated by the driving current to the inhibitory neurons. The parameter
space for this model network was much larger than for the homogeneous,
all-to-all network.  The additional parameters include the degree of
sparseness and the dispersion of driving currents across network neurons.
Therefore, we did not attempt a systematic study of the entire parameter
space, rather we report on the results obtained with two representative
parameter sets. For these sets, small changes in parameter settings did
not result in qualitative changes in the network states, indicating that
the behavior we found was robust.
The first parameter set was adjusted to obtain network states with mean
firing rates that were much less than the oscillation frequency. The
coupling constants were $(g_{ee}, g_{ei}, g_{ie},
g_{ii})=(0.1,0.5,0.3,0.5)$. The 
second parameter set was tuned such that the excitatory rates were
between 15 and 35 Hz in order to reproduce the CRFs measured by
\cite{rey00}. This was accomplished by reducing the strength of the
inhibitory synapses onto the excitatory neurons from $g_{ie}=0.3$ to 0.2. The
excitatory neurons in the network were divided into two groups (see
Methods), a hundred neurons with a high driving current (because the
stimulus matched the preferred stimulus feature of the neurons), and
three hundred neurons with a lower driving current (because stimulus did
not exactly match the neurons' preferred stimulus feature). These
neurons are referred to as the 
top-100 and bottom-300, respectively. For set 1, oscillation frequencies
between 22 and 50 Hz were obtained, whereas set 2 yielded a range between
25 and 60 Hz.

\begin{figure}
\centering
\epsfig{file=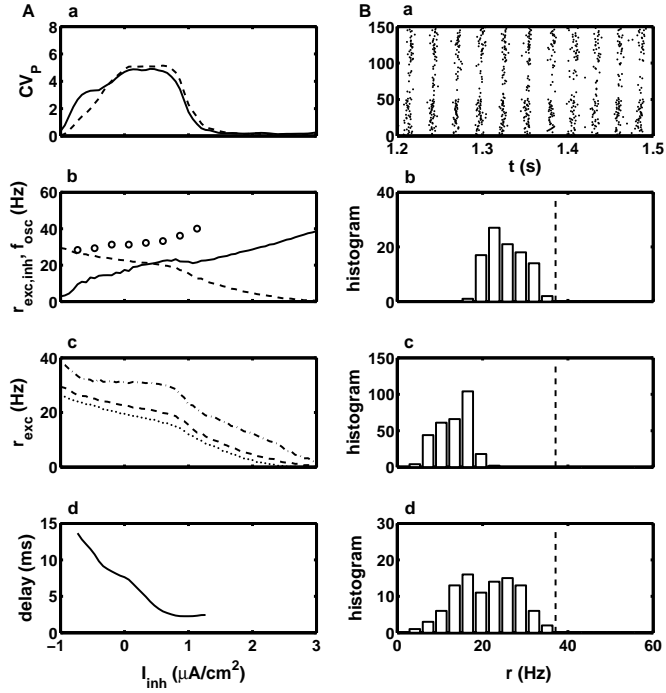, width=4in}
\caption{Locking states in sparse, heterogeneous networks. In panel A,
we plot as a function of $I_{inh}$, (a) the degree of synchrony ($CV_P$) for
excitatory (dashed line) and inhibitory neurons (solid line); (b) the
oscillation frequency (fosc, open circles), mean firing rate across all
excitatory (dashed line) and inhibitory neurons (solid line); (c) mean
firing rate across the top-100 (dot-dashed line), the bottom-300 (dotted
line) and all excitatory neurons (dashed line); (d) the delay between
inhibitory and the initial (top-100) excitatory volley.  (B) The
locked-state for $I_{inh}=0.93 \, \mu\textrm{A/cm}^2$. (a) Rastergram,
with, from top to bottom, 50 spike trains from 
the inhibitory neurons, 50 spike trains from the bottom-300 and 50 spike
trains from the top-100 excitatory neurons. Histograms of the
distribution of firing rates across network neurons for (b) the top-100,
(c) the bottom-300 excitatory neurons and (d) the inhibitory neurons. The
dashed vertical line in (b-d) indicates the oscillation frequency. Data
was obtained from a sparse, heterogeneous network with parameter set 2
with $I_{exc}=3.5 \, \mu\textrm{A/cm}^2$.}
\label{f:locksparse}
\end{figure}

The 1:1 locked states in the sparse, heterogeneous networks consisted, as
before, of an excitatory volley followed by an inhibitory volley (Figure
\ref{f:locksparse}Ba, parameter set 2). These synchronous states arose
via a robust 
version of the mechanism observed in the homogeneous, all-to-all network.
The parameters were tuned such that a synchronous volley, in which only a
fraction of the excitatory neurons participated, could elicit a
synchronous inhibitory volley. The less excitable neurons could fire
between the initial excitatory volley and the recruited inhibitory
volley. This implies that the firing rate of the bottom-300 should depend
on the value of the delay between the inhibitory and initial excitatory
volley (Figure \ref{f:locksparse}Ad). Indeed, there was a range of
$I_{inh}$ values for which 
the rate of the top-100 neurons increased (data not shown) or remained
approximately constant (Figure \ref{f:locksparse}Ac), but the rate of
the bottom-300 
(Figure \ref{f:locksparse}Ac) and the delay decreased (Figure
\ref{f:locksparse}Ad). During moderately 
synchronous states, the top-100 neurons typically had firing rates below,
but relatively close to, the oscillation frequency with a broad
dispersion ($I_{inh}=0.93$, $I_{exc}=3.5$, Figure
\ref{f:locksparse}Bb). The bottom-300 (Figure \ref{f:locksparse}Bc) 
and the inhibitory neurons (Figure \ref{f:locksparse}Bd) had rates
that were generally 
much lower than $f_{osc}$. For this particular example, they took values
between approximately 4 and 23 Hz and between 4 and 34 Hz, respectively,
with $f_{osc}$ being approximately 37 Hz. For parameter set 1, even larger
differences between the oscillation frequency and the mean firing rate
were obtained (data not shown). For highly synchronous states (example:
parameter set 2, $I_{inh}=0.27$, $I_{exc}=4.75$), a large fraction of
the top-100 
fired at the oscillation frequency, with the bottom-300 firing at a lower
rate and at a significant delay (data not shown). In that case, the
firing rate of the most depolarized excitatory neurons determined the
oscillation frequency. Thus, it is the subsystem consisting of the
top-100 excitatory neurons and the interneurons, which shows behavior
similar to that obtained in the homogeneous, all-to-all network. The main
difference is that the overall excitatory and inhibitory firing rate can
be different from the oscillation frequency \cite{bru99,bru03,gei05}
and that the states are much more robust against heterogeneity.

\begin{figure}
\centering
\epsfig{file=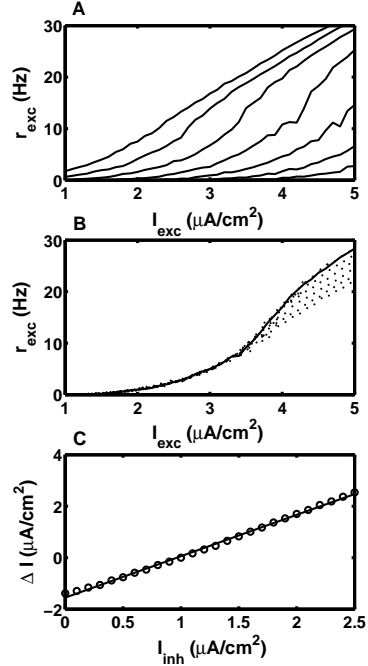, width=2in}
\caption{The inhibitory driving current modulates the sensitivity of the
$r_{exc}-I_{exc}$ curves. (A) $r_{exc}-I_{exc}$ curves for, from top to bottom,
$I_{inh}=0.05$, 0.45, 0.85, 1.25, 1.65, 2.05, and 2.45
  $\mu\textrm{A/cm}^2$. (B) The same curves, 
but now each curve is shifted over a distance $\Delta I(I_{inh})$,
chosen to make the 
curves as similar as possible to a reference curve with $I_{inh}=1.0$. (C)
Shift distance $\Delta I(I_{inh})$ as a function of the inhibitory
driving current. Data was obtained from a sparse, heterogeneous
network with parameter set 1.} 
\label{f:sensmod}
\end{figure}

We also investigated how the $r_{exc}-I_{exc}$ curves depended on the
inhibitory driving current (Figure \ref{f:sensmod}A, parameter set
1). In general, there were two regimes in the $r_{exc}-I_{exc}$
curve. For low firing rates, $r_{exc}$ varied as a 
power of $I_{exc}$, $r_{exc}=0.3(I_{exc}-B)^A$ \cite{hans02,mill02},
here A and B are fitting
parameters. A is approximately independent of $I_{inh}$, $A=3.2\pm 0.1$ , the
``shift'' B is discussed below. For higher firing rates, the rate of change of
$r_{exc}$ with $I_{exc}$ first increased, but decreased soon after
that, when the firing rate reached saturation. The low firing rate
(power law) portion 
of the curves could be made to fall on top of a reference curve
($I_{inh}=1.0 \, \mu\textrm{A/cm}^2$) by
shifting them over a distance $\Delta I$ along the $I_{exc}$
coordinate (Figure \ref{f:sensmod}B). 
This means that $B\approx \Delta I(I_{inh})+1$. The shift $\Delta I$
was a linear function of $I_{inh}$ (Figure \ref{f:sensmod}C, 
slope=1.6, $y$-intercept: -1.6 $\mu\textrm{A/cm}^2$). For parameter
set 2, the shifts were 
present for a smaller range of inhibitory current values, but the curves
overlapped for higher firing rates of up to 20 Hz (results not shown).
These results are a proof of principle that the sensitivity of the
network activity can be modulated via the inhibitory driving current.

\subsection{Contrast and attention dependence of the cortical response}

\begin{figure}
\centering
\epsfig{file=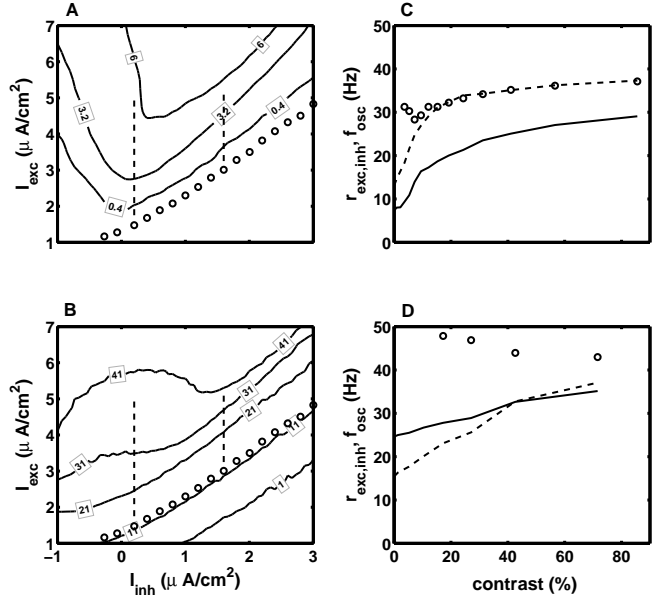, width=4in}
\caption{A-B: The trajectory in $I_{exc}-I_{inh}$ space obtained by varying
contrast. We plot the contour lines (solid lines) for (A) the inhibitory
$CV_P$ and (B) the excitatory firing rate for the top-100 neurons. In each
panel we also show the lines of varying contrast for a fixed level of
attentional modulation (dashed lines) and the line corresponding to zero
contrast and varying attentional condition (circles). C-D: In each panel,
we plot as a function of contrast, the firing rate of the top-100
excitatory (dashed line) and inhibitory neurons (solid line) as well as
the oscillation frequency (open circles), We show the model response when
attention is (C) directed into the receptive field and (D) when it is
directed away from the receptive field. Data was obtained from a sparse,
heterogeneous network with parameter set 2.} 
\label{f:ispace}
\end{figure}

The contrast-response function measured in experiments is represented by
the firing rate of the excitatory neurons in the network. Experimental
measurements of the CRF in LGN and V1 can be fit by the relationship
\cite{albr82,sclar82,ali04}:

\[ f(c)=f_m \frac{c^n}{c_0^n+c^n}
\]
Here $f_m$ is the firing rate at saturation, $c$ is the stimulus
contrast, $c_0$ 
is the contrast for which the firing rate is half the saturation rate,
and $n$ can be interpreted as a steepness parameter. Strong contrast
saturation effects are already present in neurons in LGN, V1 and V2, that
provide direct or indirect inputs to V4. Therefore, we do not have to
fully account for saturation effects in the network model; rather, we
assume that the driving currents, which represent the feed forward
inputs, vary with contrast according to $f(c)$. We took an $f(c)$ appropriate
for LGN inputs from \cite{troy98}: $c_0=13.3\%$, $n=1.2$,
$f_m=53$ Hz. As 
an additional simplification, we assume that the driving currents are a
linear combination of the CRF $f(c)$ and the attentional state $a$. That is:

\[ \left[ \begin{array}{c} I_{exc}(c,a) \\ I_{inh}(c,a) \end{array}
  \right] = \left[ \begin{array}{c} I_{exc0} \\ I_{inh0} \end{array}
  \right] + \left[ \begin{array}{cc} A_{11} & A_{12} \\ A_{21} & A_{22}
  \end{array} \right] \left[ \begin{array}{c} f(c) \\ a \end{array} \right]
\]
Presently, the coefficients in the matrix are not known to sufficient
detail. For instance, we do not know what is the relative increase in
drive to excitatory neurons with contrast compared with the increase in
drive to inhibitory neurons. By the same token, we do not know whether
attention, which could potentially be mediated by either cholinergic or
glutamatergic projections \cite{hass04,coull05,mils05,sart05}, affects
the interneurons 
more strongly than the excitatory neurons. These coefficients were
therefore treated as free parameters. There is also a baseline current,
$I_{exc0}$ and $I_{inh0}$, in order to get excitatory rates close to
15 Hz for zero 
contrast (corresponding to a gray screen \cite{rey00}). We
report results based on one choice of parameter values: $I_{exc0}=3.007$,
$I_{inh0}=1.600$, $A_{11}=0.0898$, $A_{12}=-1.532$, $A_{21}=0$,
$A_{22}=-1.400$. $A_{21}/A_{11}$ is the 
ratio between contrast-induced current to interneurons over that to
excitatory cells. In the present case it is zero, implying that
interneurons are not affected by the amount of contrast. However, similar
results were found for ratios between -0.15 and 0.20.

The network responses were calculated on a two-dimensional grid of $I_{inh}$
and $I_{exc}$ values. We used data obtained from simulations of the sparse,
heterogeneous network with parameter set 2. The firing rate was averaged
across the top-100 neurons, since these were the neurons most strongly
activated by the stimulus.  Then we constructed the responses as a
function of $c$ for a given value of
``attentional modulation'', $a$. Here the ``attentional modulation''
 could take only two values: $a=0$ for the attend-away state, and $a=1$ for
the attend-into-RF state. As $c$ is varied, a line is traced out in
$I_{exc}-I_{inh}$ space. The rate at which the line is traced out with contrast
is not uniform because of the saturation, $f(c)$. In Figure
\ref{f:ispace}A and B, we 
show the line for the attend-away condition and to the left of it (at
lower $I_{inh}$ values), the line for when attention is directed into
the neuron's receptive field. The starting point of each line
corresponds to zero contrast for the particular attentional
condition. The starting points 
lie on a line characterized by constant $c$ and varying $a$. There have been
experiments in which attention increased the spontaneous activity,
whereas in other experiments no significant changes were observed
\cite{moran85,luck97}. Here,
the parameters of this line were chosen so that the firing rate at zero
contrast is approximately independent of attentional state.

\begin{figure}
\centering
\epsfig{file=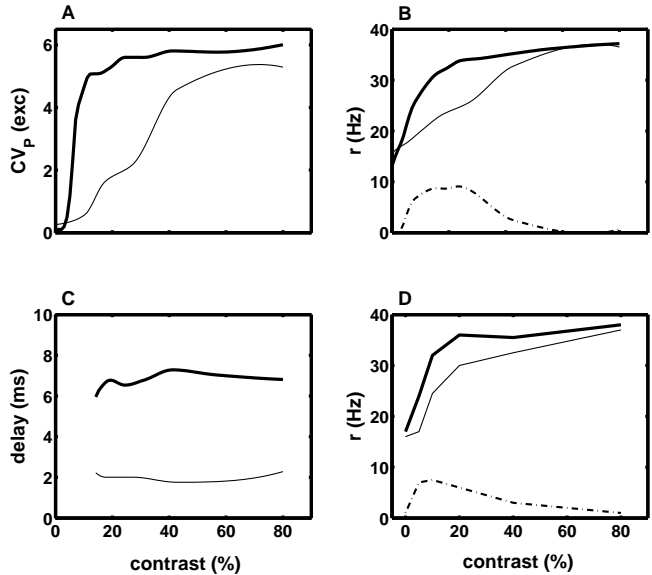, width=4in}
\caption{Comparison of the response of the top-100 excitatory neurons
between the attend-into-the-receptive-field (thick solid lines) and the
attend-away condition (thin lines). We plot as a function of contrast,
(A) the $CV_P$, (B) the firing rate, (C) the delay between inhibition and
excitation. In (D) we reproduced the experimental data shown in Figure 5
of (Reynolds et al., 2000). Simulation data were the same as in Figure
\ref{f:ispace}. 
}
\label{f:contrast}
\end{figure}
For weak contrast, in the attend-away condition, interneurons fired at
about 25 Hz, whereas the excitatory neurons had a much lower firing rate
(Figure \ref{f:ispace}D). As the contrast was increased, the
excitatory rate increased 
significantly, whereas the inhibitory rate increased at a much lower
rate. At approximately 50\% contrast, the excitatory and inhibitory rate
became quite similar, and the degree of synchrony in the gamma-frequency
range increased (Figure \ref{f:contrast}A).  The delay between
inhibition and excitation 
remained approximately constant during this manipulation (Figure
\ref{f:contrast}C). The
effect of attention was modeled as a decrease in driving current to the
inhibitory neurons. With attention, the excitatory firing rate increased
much faster with increasing contrast and saturated for lower values of
contrast (approximately 20\%, Figure \ref{f:ispace}D), synchrony was
obtained at a 
lower contrast as well (Figure \ref{f:contrast}A). The degree of synchrony, as
quantified using the $CV_P$, was significantly higher with attention for
contrast values up to 60\% (Figure \ref{f:contrast}A). Even for higher
contrast, when the 
firing rates for both conditions were approximately the same, there still
was a small difference in $CV_P$ values. During the synchronous state, with
attention into the receptive field, the delay between excitation and
inhibition was higher than in the attend-away condition (Figure
\ref{f:contrast}C). The 
magnitude of the attentional modulation of firing rate in the model
(Figure \ref{f:contrast}B) is similar to that obtained in the
experiments by \cite{rey00} which were reproduced in Figure
\ref{f:contrast}D. For the experimental as 
well as the computational results, the saturation firing rate was not
affected by the attentional condition, indicating that the effect of
attention is more consistent with a shift in sensitivity than a change of
gain. However, the CRF curves themselves are not exactly shifted versions
of each other. In the model, we obtained shifts in sensitivity by
modulating the inhibitory driving current (Figure \ref{f:sensmod}).

\subsection{Delay between inhibition and excitation affects downstream
  neurons.} 

\begin{figure}
\centering
\epsfig{file=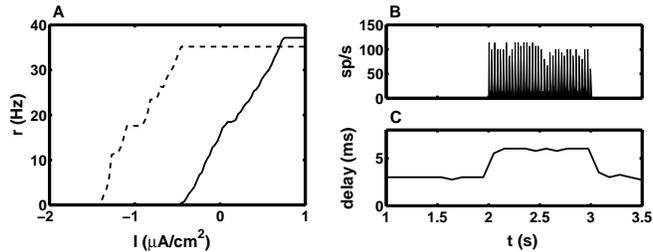, width=4in}
\caption{Downstream neurons were sensitive to the delay between
inhibition and excitation. (A) We plot the firing rate of a downstream
neuron for a small delay between inhibition and excitation ($\Delta
t=3.01$ ms, solid line) and a larger delay ($\Delta t=6.03$ ms, dashed
line). (B) When the delay 
was dynamically switched from the low value (at $t<2000$ ms) to the higher
value (during the interval $2000<t<3000$ ms), the neuron's firing rate
increased instantaneously. We plot the spike time histogram 
for the responses of a single downstream neuron across 20 trials. (C)
Time evolution of the delay for the network used as input in (B).} 
\label{f:delaymod}
\end{figure}
We also determined how changes in network state might affect downstream
neurons. For the present investigation, we assumed that the downstream
neuron received excitatory and inhibitory inputs from the network. This
means that it is local, because connections between different cortical
areas are predominantly excitatory \cite{salin95}. We are
considering the effect of the vertical projection from layer 4 to layer
2/3 and from layer 2/3 to layer 5, rather than the long-range horizontal
projections that link different microcolumns \cite{call98,doug04} 
The downstream neuron was modeled in the same way as an excitatory neuron
in the network, with a driving current $I=-0.5$ (in $\mu\textrm{A/cm}^2$),
a total excitatory input conductance of 0.6 (in $\textrm{mS/cm}^2$,
400 inputs) and a total inhibitory 
conductance of 0.15 (100 inputs). We ran the homogeneous, all-to-all
network for an extended period of time for two different values of the
inhibitory driving current (parameters as in Figure \ref{f:inhidep},
with $I_{exc}=3.94$, 
$I_{inh}=0.72$ and 0.28, respectively). For both values the network was
synchronized. In addition, for both states, the oscillation frequency,
the excitatory firing rate and inhibitory firing rate were also
approximately equal. Only the delay was significantly different.
The firing rate of the downstream neurons was increased strongly when the
network delay was increased (Figure \ref{f:delaymod}A,B). The firing
rate versus driving 
current curve was shifted to the left when the delay was increased
(Figure \ref{f:delaymod}A). Hence, increasing the delay made the
neuron more sensitive, but it did not increase the saturation firing rate.

\section{Discussion}

\subsection{Summary of the simulation results}

{\it Synchrony changes are obtained by modulating the driving current to
interneuron networks.} We studied firing rate modulation of a strongly
coupled model network, which was representative of networks in the
superficial layers of the cortex \cite{doug04,yoshi05}. Our
simulations of the homogeneous, all-to-all network 
revealed that increasing the drive to excitatory neurons (Figure
\ref{f:excidep}) had a 
different effect than decreasing the drive to inhibitory neurons (Figure
\ref{f:inhidep}). In the former case, the firing rate of excitatory
and inhibitory 
neurons increased, typically with little changes in the delay between
inhibition and excitation and in the degree of synchrony. By contrast,
when the drive to inhibitory neurons was decreased, their rate decreased,
but the rate of the excitatory neurons increased. When the inhibitory
neurons fired at a higher rate than the excitatory neurons, the initial
difference in firing rate decreased with decreasing interneuron drive,
leading to an increase in synchrony by virtue of a locking phenomenon.
Decreasing interneuron drive when the network was already synchronous led
to smaller changes in the rate, but the delay between inhibition and
excitation increased instead. These results were robust against the
effects of noise (Figure \ref{f:hnoise}), sparse synaptic connectivity and
heterogeneity (Figure \ref{f:locksparse}). For sparse, heterogeneous
networks, the mean 
excitatory and inhibitory rates during 1:1 locking states were not
necessarily equal to the oscillation frequency.
In our simulations (Figure \ref{f:inhidep}Ac, \ref{f:hnoise}B and
\ref{f:locksparse}Ab), during 1:1 locking, the 
oscillation frequency increased, albeit moderately, with increasing
inhibitory driving current. This is not necessarily the only possibility.
In other operating regions of the network, an increase in interneuron
activity could lower the firing rate of the excitatory neurons, which in
turn could lower the oscillation frequency.

{\it Attentional modulation of the contrast response function.} The preceding
results on the modulation of synchrony and firing rate with the driving
current to inhibitory and excitatory neurons were used to infer how the
effects of attention could be modeled in the cortical circuit. We made
the following assumptions to account for attentional modulation of the
CRF. First, contrast mostly activates the excitatory neurons and to a
lesser extent inhibitory neurons. Second, that interneurons fired at a
higher rate than excitatory neurons. Third, that attention led to a
reduction in the drive to the interneurons. This last assumption may seem
counterintuitive and is discussed in the section ``Model
assumptions''. Given the preceding assumptions, the model reproduces
the following experimental observations (Figure \ref{f:ispace} and
\ref{f:contrast}). (1) Contrast leads to 
increases in the rates of excitatory and inhibitory cells, but does not
significantly change the delay between inhibition and excitation. (2)
Synchrony is only present for high contrast stimuli. (3) Attention
increases the rate of excitatory cells, but decreases the rate of
inhibitory cells. (4) Attention shifts the onset of synchrony to lower
contrasts. Furthermore, it predicts that when there already is synchrony,
attention increases the delay between inhibition and excitation. We do
not claim that modulation of interneuron drive is the only way of
shifting CRFs. For instance, increasing the drive to excitatory neurons
or increasing the efficacy of recurrent inhibitory synapses could also
lead to modulation of CRFs. We have not evaluated these possibilities
within the context of the present model.

{\it Changes in the delay between inhibition and excitation could modulate the
firing rate of downstream neurons.} We determined how changes in network
state might affect downstream neurons (Figure \ref{f:delaymod}). For
the present 
investigation, we assumed that the downstream neuron received excitatory
and inhibitory inputs from the network. Our results indicate that the
changes in delay obtained from our network model could switch downstream
neurons from a non-responsive to a responsive state. The response of
neurons that were already responsive changed to a lesser extent. Hence,
for the parameter values used here, the main effect was to increase the
number of downstream neurons that are responsive to the stimulus. The
changes in delay were obtained with small changes in firing rate
hence, the strongest attentional modulation of firing rate may occur
downstream from where attention actually modulated the driving currents
to neurons.

\section{Model assumptions}

{\it Representing synaptic inputs by currents.} The feed-forward
feature-selective and top-down modulatory inputs, as well as inputs from
other neurons in the same layer but outside the network, were modeled as
a time-varying current. The replacement of synaptic inputs that elicit
conductance changes together with a driving current by a pure current
without conductance changes is an approximation. There is disagreement
about whether it is possible to get exactly the same spike train
statistics with a time-varying current compared with those that are
obtained for a conductance drive
\cite{rau03,rud03,rich04}. Nevertheless, the mean and the 
variance of a current drive affect in different ways the firing rate
response of an isolated neuron \cite{ties00,cha02,fell03}
 and the coherence of a network \cite{ties00net}. When the input rate
 of synaptic inputs is increased, it 
increases the mean as well as the variance of the current. Here we used a
current drive, so that we could independently vary the mean and variance
in the simplest possible way.  When the mean and variance were co-varied
to represent increasing the rate of synaptic inputs, synchrony and delay
modulations similar to those reported here were obtained (results not
shown).

{\it Can attention reduce the drive to the interneurons?} Attention has been
associated with activation of cholinergic projections
\cite{hass04,coull05,mils05,sart05}.  Cholinergic effects mediated
through the M1 receptor are 
excitatory, because they inhibit a voltage-sensitive K+ channel. In
contrast, effects mediated through the M2 and M4 receptor are inhibitory
because acetylcholine activates an inward rectifying K+channel
\cite{feld97}. The M1, M2 and M4 receptors are all expressed in the
neocortex \cite{feld97}. However, in the neocortex,
acetylcholine is generally thought to have excitatory effects on neurons
\cite{krn93}. How can it cause the drive to the interneurons to
decrease? There are at least two possible mechanisms. There could be a
second set of interneurons that is activated by attention, these neurons
might inhibit the interneurons studied here, thus reducing their
rate \cite{wang04}. There are multiple, functionally distinct networks 
of interneurons in cortex that could play this role
\cite{bei03,mark04}. Cortical interneurons have been classified 
based either on their physiological characteristics, or on their
morphological characteristics or sometimes based on their neurochemical
characteristics. The link between the results of different classification
methods has not been conclusively established. We speculate that the
second set of neurons may either consist of interneurons that are
physiologically classified as low threshold spiking (LTS), since these
are more sensitive to neuromodulators than fast spiking interneurons
\cite{bei03}, or those that are morphologically classified as
interneuron targeting cells (and contain the calcium binding protein
calbindin) \cite{mark04,wang04}.
The levels of muscarinic receptor expression varies across brain areas,
M1 is expressed at the highest level in the neocortex, whereas M2 is
expressed at the highest level in the cerebellum \cite{feld97}.
This would seem to suggest that the expression of receptors can be
tightly controlled, hence, that it could be cell-type specific, with one
type of interneuron more sensitive to a specific neuromodulator than the
other. For instance, the interneurons in the model network could be
inhibited, or even not affected at all, by acetylcholine, whereas the
postulated second set of interneurons are excited by acetylcholine.
Another possibility, suggested by measurements on recurrent excitatory
synapses in the hippocampus \cite{hass95}, is that
acetylcholine reduces the efficacy of excitatory synapses onto inhibitory
cells. The synaptic strength is modulated by acetylcholine via
autoreceptors on the presynaptic neuron \cite{coop96}. Hence, if
excitatory synapses onto excitatory neurons are modulated by
acetylcholine \cite{hass95,hass04,sart05}, it could also be the case
that those onto inhibitory neurons are similarly modulated. The
converse could also be true, since 
it was shown that inhibitory synapses could have different facilitation
and depression properties based on their postsynaptic target
\cite{gupta00}. Hence, this issue needs to be resolved
experimentally.

\subsection{Comparison to previous theoretical studies.}

Previous theoretical studies and model simulations have shown that
interneuron networks can be made to synchronize in the gamma-frequency
range \cite{whit95,wang96,golomb00,ties00net,bart01,aradi02,bart02,oluf03}.
The synchronization is robust against heterogeneity in the 
physiological properties of the interneurons \cite{wang96,white98},
noisy background activity \cite{ties00net}
and sparse connectivity \cite{wang96,golomb00,borg03}. In other
investigations it was found that 
there are two mechanisms for obtaining gamma frequency oscillations in
mixed excitatory-inhibitory networks
\cite{bush96,ties01,borg03,bru03,hans03,borg05,borg05b}. In the first 
mechanism, the sparsely firing excitatory cells are entrained to the
periodic inhibition produced by the synchronized interneuron network.
This mechanism is referred to as interneuron gamma (ING)
\cite{whit00}. In the second mechanism, the activity of excitatory
neurons 
recruits interneurons, which in turn temporarily shuts off the excitatory
neurons, after which the whole process repeats itself. This mechanism is
known as pyramidal-interneuron gamma (PING) \cite{whit00}.
The focus of the present investigation differs from the preceding ones.
We address two issues. First, is it possible to vary the degree of
synchrony without significantly affecting the firing rate of excitatory
and inhibitory neurons? Second, can the degree of synchrony be modulated
on rapid time scales? Previously, we addressed these issues for an
isolated interneuron network \cite{ties04}. Here, we
find that the degree of synchrony and the delay between inhibition and
excitation, can be modulated dynamically in a mixed excitatory-inhibitory
network with only minor changes in the neurons' firing rate. The model
used in this paper and the cited papers are far 
too simple to account for the detailed dynamics of a cortical column. Our
goal is to use the present model as a part of a much larger - 1000 or
more columns - model of the visual cortex. Therefore we had to make some
simplifications to make the calculations feasible. A different line of
research is to include more ionic channels, use multicompartmental single
neuron models and account for more distinct types of neurons. A recent
example is the work by Traub and coworkers \cite{traub05}.

{\it Modulation of delay between inhibition and excitation and models for
stimulus competition.} When two visual stimuli are presented
simultaneously, the corresponding neural activity patterns may compete
for control of neurons in downstream areas \cite{des95}.
In a recent paper it was argued that synchronous inhibition could help
with this stimulus competition \cite{borg05b}. The authors
proposed that a strong stimulus (the winner) would spike the responding
neurons before the synchronous inhibition arrives and the neurons
responding to the weak stimulus (the loser) would be prevented from
spiking by the inhibition. In this way the inhibition only affects the
activity of the loser. By contrast, tonic inhibition would affect both
the winner and loser equally. Here we propose that attention can increase
the delay between inhibition and excitation: this would make it harder to
stop a weak stimulus that spikes neurons immediately after the spikes of
neurons responding to the strong stimulus. The two models are not in
conflict because they speak to different situations. First, our model is
meant to represent one cortical column, rather than multiple columns that
compete. Second, the downstream effects in our model are vertical, from
layer 4 to layer 2/3 and from layer 2/3 to layer 5, rather than
horizontal. Further theoretical studies are necessary to properly
integrate the vertical and horizontal components of attention.

\subsection{Comparison to experimental results and future work}

{\it Response versus contrast gain.} The model could account for shifts in the
CRF with attention observed for V4 neurons by
\cite{rey00,rey03}. Their results provide support for a 
contrast gain model rather than a response gain model. However, in other
experiments \cite{mcad99}, where a neuron's orientation
tuning curve was measured, it was concluded that the 
response gain model was a more appropriate description. These two
experiments are not directly comparable because the behavioral task was
different and the responses in \cite{mcad99} may not have
reached contrast-saturation. In addition, varying stimulus orientation
alters the identity of the cortical neurons that drive the neuron,
whereas contrast does not, rather it increases the rate of the neurons
providing input. As a result, properties, such as neural synchrony, are
modulated differently by contrast compared with orientation
\cite{kohn05}. Further experimental studies of the modulation of firing
rate with attention, luminance contrast and the value of stimulus
features are needed \cite{rey04}.

{\it Synchrony modulation and contrast.} In experiments conducted to measure
attentional modulation of synchrony, the stimulus is presented for a long
time \cite{fries01,tay05}, on the order of seconds,
whereas contrast-response curves are usually obtained using short 50-250
ms stimuli \cite{rey03}. Hence, it may not be
appropriate to directly compare these experimental results. To the best
of our knowledge, there are no experiments in which the attentional
modulation of the synchrony-versus-contrast curves was measured directly.
Previous studies did, however, address the relation between
gamma-frequency oscillations, stimulus orientation and contrast. Here we
briefly summarize these studies. Gamma-frequency oscillations could be
measured in recordings from feline and primate primary visual cortex when
the stimulus was close to optimal in terms of the cell's preferred
orientation, it had a high contrast and it was large \cite{gray97}. In
macaque V1 it was found that the spectral 
content of the local field potential in the gamma-frequency range
increased with contrast \cite{hen05}. The oscillatory
synchrony is stimulus specific and is absent or very weak during
spontaneous activity \cite{gray97}. The gamma
oscillations were localized on the cortical surface; only those areas
that responded to the region of visual field where the stimulus was
located were synchronized \cite{rols01}. The degree of synchrony
was tuned for stimulus orientation and it increased with contrast
\cite{gray97}. Furthermore, frequency bands of the LFP power
spectrum were tuned for orientation, and spatial and temporal frequency
\cite{kay04}. Even the oscillation frequency itself could
depend on stimulus features. For instance, when the speed of a moving
stimulus bar was increased, the oscillation frequency also increased
\cite{gray97}. The preceding results show that
synchrony and rate modulations produced by the model could be feasible.
However, further experiments are needed to determine how the
synchrony-versus-contrast curves are modulated by attention. Our
simulation results predict that there should be a group of interneurons
that lower their rate with attention. It is therefore key to record from
identified cortical interneurons in awake animals, which is challenging
from an experimental point of view \cite{con02,swa03}.

\acknowledgements

We thank Jean-Marc Fellous and Vincent Toups for comments on the
manuscript. This research was supported by startup funds provided by the
University of North Carolina at Chapel Hill.

\appendix

\section{The model equations}

{\it Interneuron model}

The interneuron was modeled as a single compartment with
Hodgkin-Huxley-type voltage-gated sodium and potassium currents and a
passive leak current \cite{wang96}. The equation for the membrane
potential of the model neuron is:

\[C_m \frac{\mathrm{d} V}{\mathrm{d} t}=-I_{Na}-I_K-I_L-I_{GABA}
-I_{AMPA}+I+C_m\xi
\]
where $I_L=g_L(V-E_L)$ is the leak current,
$I_{Na}=g_{Na}m^3_{\infty}h(V-E_{Na})$ is the sodium current,
$I_K=g_Kn^4(V-E_K)$ is the potassium current, $I_{GABA}$ is the
inhibitory synaptic current and $I_{AMPA}$ is the excitatory synaptic
current. The Gaussian noise variable is denoted with $\xi$ while $I$
is the tonic drive. The gating variables are given in terms of 
$m$, $n$, and $h$ and they satisfy the equation

\[\frac{\mathrm{d} x}{\mathrm{d} t}=\zeta(\alpha_x(1-x)-\beta_x x)
\]
Here the label $x$ stands for the kinetic variable, and
$\zeta=5$ is a dimensionless time scale that can be used to tune the
temperature dependent speed with which the channels open or close. The
rate constants are:

\begin{eqnarray*}
\alpha_m = \frac{-0.1(V+35)}{\exp (-0.1(V+35))-1}, \quad & 
\beta_m =& 4\exp (-(V+60)/18) \\
\alpha_h = 0.07\exp (-(V+58)/20), \quad & 
\beta_h =& \frac{1}{\exp (-0.1(V+28))+1} \\
\alpha_n = \frac{-0.01(V+34)}{\exp (-0.1(V+34))-1}, \quad & 
\beta_n =& 0.125\exp (-(V+44)/80) \\ 
\nonumber
\end{eqnarray*}
and the asymptotic values of the gating variables are:

\[x_{\infty}(V)=\frac{\alpha_x}{\alpha_x+\beta_x}
\]
where $x$ stands for $m$, $n$, and $h$.
We made the approximation that $m$ follows the asymptotic value
$m_{\infty}(V)$ instantaneously \cite{wang96}. 

\pagebreak

\noindent{\it Pyramidal neuron model}

A different single compartment Hodgkin-Huxley-type model was used for the
dynamics of the pyramidal neuron \cite{golomb97}. The membrane
potential obeys the following differential equation:

\begin{eqnarray*}
C_m \frac{\mathrm{d} V}{\mathrm{d} t}&=&-I_{Na}-I_{NaP}-I_K-I_{Kdr}-I_{KA}
-I_{Kslow}\\
 & & -I_L-I_{GABA}-I_{AMPA}+I+C_m\xi\\
\nonumber
\end{eqnarray*}
where $I_L=g_L(V-E_L)$ is the leak current,
$I_{Na}=g_{Na}m^3_{\infty}h(V-E_{Na})$ is the sodium current,
$I_{NaP}=g_{NaP}p_{\infty}(V-E_{Na})$ is the persistent 
sodium current, $I_{Kdr}=g_{Kdr}n^4(V-E_K)$ is the delayed rectifier
potassium current, $I_{KA}=g_{KA}a^3_{\infty}b(V-E_K)$  is the 
A-type potassium current, $I_{Kslow}=g_{Kslow}z(V-E_K)$
is the slow potassium current, $I_{GABA}$ is the 
inhibitory synaptic current and $I_{AMPA}$ is the excitatory synaptic current.
The Gaussian noise variable is denoted by $\xi$ while $I$ is the tonic
drive. The kinetic variables $h$, $n$, $b$ and $z$ satisfy the equation

\[\frac{\mathrm{d} x}{\mathrm{d} t}=\frac{x_{\infty}(V)-x}{\tau_x}
\]
The rate constants are:

\begin{eqnarray*}
m_{\infty}&=&[\exp (-(V+30)/9.5)+1]^{-1} \\
h_{\infty}&=&[\exp ((V+53)/7)+1]^{-1} \\
\tau_h&=&0.37+2.78\times [\exp ((V+40.5)/6)+1]^{-1} \\
p_{\infty}&=&[\exp (-(V+40)/5)+1]^{-1} \\
n_{\infty}&=&[\exp (-(V+30)/10)+1]^{-1} \\
\tau_n&=& 0.37+1.85\times [\exp ((V+27)/15)+1]^{-1} \\
a_{\infty}&=&[\exp (-(V+50)/20)+1]^{-1} \\
b_{\infty}&=&[\exp ((V+80)/6)+1]^{-1} \\
\tau_b&=& 15 \, \textrm{ms} \\
z_{\infty}&=&[\exp ((V+39)/5)+1]^{-1} \\
\tau_z&=& 75 \, \textrm{ms} \\
\nonumber
\end{eqnarray*}
The fast gating variables $m$, $p$ and $a$ instantaneously followed their
asymptotic value $m_{\infty}$, $p_{\infty}$ and $a_{\infty}$, respectively.
The standard set of parameter values used in this paper is:

\begin{tabular}{|l|c|c|}
\hline
 Parameter (units) & Pyramidal Neurons & Interneurons \\
\hline
$E_L$ (mV) & -70   &  -65 \\
\hline
$E_{Na}$ (mV)   &     55   &   55 \\
\hline
$E_K$ (mV) & -90  &   -90 \\
\hline
$E_{AMPA}$ (mV) &     0   &    0 \\
\hline
$E_{GABA}$ (mV)  &    -75  &   -75 \\
\hline
$g_L$ (mS/cm$^2$)  &   0.02 &   0.1 \\
\hline
$g_{Na}$ (mS/cm$^2$) &   24  &    35 \\
\hline
$g_{NaP}$ (mS/cm$^2$) &  0.07 &   - \\
\hline
$g_{Kdr}$ (mS/cm$^2$) &  3   &    - \\
\hline
$g_{KA}$ (mS/cm$^2$) &   1.4 &    - \\
\hline
$g_{Kslow}$ (mS/cm$^2$)   &     1   &    - \\
\hline
$g_K$ (mS/cm$^2$)  &   -  &     9 \\
\hline
$C_m$ ($\mu$F/cm$^2$) &  1   &    1 \\
\hline
\end{tabular}

\noindent The initial values of the membrane potential at the beginning of the
simulation were chosen randomly from a uniform distribution between -70
mV and -50 mV. The gating variables were set to their asymptotic
stationary values, $x_{\infty}$, corresponding to the starting value, 
$V$, of the membrane potential.

The tonic drive, $I$, for excitatory (inhibitory) neurons, was the sum of a
common component $I_{exc}$ ($I_{inh}$) and a heterogeneous component
that varied across neurons with a variance
$\sigma^2_{exc}$ ($\sigma^2_{inh}$). The common component was varied
on a two-dimensional grid of current values in the range
$-2<I_{inh}<1$ $\mu$A/cm$^2$ and $1<I_{exc}<6$ $\mu$A/cm$^2$.
The Gaussian synaptic noise $\xi_i$ in the current of neuron $i$ is
chosen such that $\langle \xi_i(t)\rangle =0$ and $\langle
\xi_i(t)\xi_j(t')\rangle =2\lambda \delta (t-t') \delta _{ij}$. 
On each integration time step, the noise was drawn independently
from a distribution between $-\sqrt{6\lambda/dt}$ and $\sqrt{6\lambda/dt}$,
where $dt$ was the time step. The differential equations are integrated using a
second-order Runge-Kutta method with a time step of $dt=0.01$ ms or
0.05 ms \cite{ger99,press92}.

\pagebreak
\noindent{\it Synaptic models}

Each synapse is modeled using a gating variable $s_{ji}$, with $j$ being the
index of the postsynaptic neuron and $i$ being the index of the presynaptic
neuron. Since there are no synaptic delays or synaptic failures in our
model, the gating variables for synapses originating from the same
presynaptic neuron will have the same value: $s_{ji}=s_i$ . For the
all-to-all network an additional simplification is possible, because
the sum across gating variables of all the synapses impinging on
postsynaptic neuron $j$ is the same for each neuron:

\[ \sum_{i(j)}s_{ji}=\sum_i s_i \equiv s_{tot}
\]
Here $i(j)$ stands for all neurons $i$ that project to neuron $j$. Hence, we
need to calculate it only once for all the excitatory and once for all
the inhibitory synapses. This resulted in a significant speed up of the
computation. $s_i^{inh}$ and $s_i^{exc}$ are the gating variables for
the inhibitory and excitatory synapses, respectively. The synapses
were labeled by the presynaptic neuron $i$. The synaptic gating
variables obey the following equation:

\begin{eqnarray}
\frac{\mathrm{d} s_i^{exc}}{\mathrm{d} t}&=&\frac{1}{\tau^{exc}}(k^{exc}\sum_k
  \delta(t-t_k^i)-s^{exc}) \label{e:s1} \\
\frac{\mathrm{d} s_i^{inh}}{\mathrm{d} t}&=&\frac{1}{\tau^{inh}}(k^{inh}\sum_k
  \delta(t-t_k^i)-s^{inh}) \label{e:s2}
\end{eqnarray}
where $t_k^i$ is the $k^{th}$ spike time generated by the $i^{th}$ neuron,
$\tau^{exc}=5$ ms, $\tau^{inh}= 10$ ms are the synaptic decay times
for excitatory and inhibitory synapses, respectively. For the sparse,
heterogenous network we used $\tau^{exc}=2$ ms, $\tau^{inh}= 8$ ms
instead. We used a sum of Dirac delta functions to account 
for the arrival of the spikes to the postsynaptic neuron, because it was
computationally less expensive. In \cite{golomb97,wang96} the
equation for the gating variable is written in the form: 

\begin{equation}
\frac{\mathrm{d} s}{\mathrm{d} t}=\alpha F(V_{pre})(1-s)-\beta s
\label{e:g}
\end{equation}
where $F$ represents the normalized concentration of the postsynaptic
transmitter-receptor complex and is assumed to be an instantaneous
sigmoid function of the presynaptic membrane potential $V_{pre}$ :

\begin{equation} 
F(V_{pre})=\frac{1}{1+\exp (-(V_{pre}-\theta_s)/\sigma_s)}
\label{e:sig}
\end{equation}
$\theta_s$ is the voltage threshold for synaptic release and takes the
value 0 mV for interneurons and -20 mV for pyramidal neurons, and
$\sigma_s=2$ mV. The constants $k_{exc}=0.44$ and $k_{inh}=1$ in our
formulation were determined by fitting the result of eqns (\ref{e:s1}) and
(\ref{e:s2}) to the result produced by (\ref{e:g}) and (\ref{e:sig}).
The synaptic currents take the values:

\begin{eqnarray*}
I_{GABA}^{exc}&=&\frac{g_{ie}}{N_{ie}}s_{tot}^{inh}(t)(V-E_{GABA}) \\
I_{AMPA}^{exc}&=&\frac{g_{ee}}{N_{ee}}s_{tot}^{exc}(t)(V-E_{AMPA}) \\
I_{GABA}^{inh}&=&\frac{g_{ii}}{N_{ii}}s_{tot}^{inh}(t)(V-E_{GABA}) \\
I_{AMPA}^{inh}&=&\frac{g_{ei}}{N_{ei}}s_{tot}^{exc}(t)(V-E_{AMPA}) \\
\nonumber
\end{eqnarray*}
where $g_{ee}$ is the conductance of the excitatory synapses onto the
excitatory neurons, $g_{ie}$ is the conductance of the inhibitory synapses
onto the excitatory neurons, $g_{ei}$ is the conductance of the excitatory
synapses onto the inhibitory neurons and $g_{ii}$ is the conductance of
inhibitory synapses onto the inhibitory neurons. $N_{ie}$ is the
number of inhibitory synapses on excitatory neurons, the quantities
$N_{ei}$, $N_{ii}$, $N_{ee}$ are defined analogously. For the
all-to-all network, $N_{ie}=N_{ii}=N_{inh}$ and
$N_{ei}=N_{ee}=N_{exc}$, here $N_{exc}$ is the number of 
excitatory neurons and $N_{inh}$ is the number of inhibitory
neurons. For the sparsely connected network, $N_{ie}=0.2 N_{inh}$,
$N_{ii}=N_{inh}$, $N_{ei}=N_{ee}=0.2 N_{exc}$. In this case $s_{tot}$ is
the sum over all the presynaptic neurons that provide input to the
postsynaptic neuron, hence, its value does depend on the identity of
the postsynaptic neuron. The actual coupling parameters are listed in
the Methods. 

\theendnotes

\bibliographystyle{klunamed}

\bibliography{jcns.bib}

\end{article}
\end{document}